# The Exact Boundary Condition to Solve the Schrödinger Equation of Many Electron System

Rajendra Prasad

Village + Post: Kamalpur, District: Chandauli, Uttar Pradesh, PIN: 232106, India

E-mail: Theochem@gmail.com

In an attempt to bypass the sign problem in quantum Monte Carlo simulation of electronic systems within the framework of fixed node approach, we derive the exclusion principle "Two electrons can't be at the same external isopotential surface simultaneously" using the first postulate of quantum mechanics. We propose the exact Coulomb-Exchange nodal surface i.e. the exact boundary condition to solve the non-relativistic Schrödinger equation for the non-degenerate ground state of atoms and molecules. This boundary condition was applied to compute the ground state energies of N, Ne, $Li_2$, $Be_2$, $B_2$, $C_2$, $N_2$, $O_2$, $F_2$, and $H_2O$ systems using diffusion Monte Carlo method. The ground state energies thus obtained agree well with the exact estimate of non-relativistic energies.

## INTRODUCTION

An ideal target of a quantum chemist/physicist is to solve the non-relativistic Schrödinger equation exactly as it describes much of the world of chemistry. If we can solve this equation at a realistic cost, we can make very precise predictions. At present, only the full-CI method is available for obtaining the exact wave function within a given basis set, but this method is too demanding computationally and therefore not affordable even for a small system.

In recent years increasing attention has been drawn to the random walk approach called diffusion Monte Carlo (DMC) method[1,2,3,4] for solving Schrödinger equation. The attractiveness of DMC method lies in that it can treat many body problems exactly. The DMC method is a projection method based on the combination of the imaginary time Schrödinger equation, generalized stochastic diffusion process, and Monte Carlo integration. The solution, it yields has only statistical error, which can be properly estimated and in principle, made as small as desired. Since in the DMC method the wave function has to be a population density, therefore, the DMC method can only describe the constant sign solution of the Schrödinger equation. This poses a serious problem if one is interested in the solution of a many electron system where the wave function is known to be antisymmetric (i.e. both positive and negative) with respect to interchange of two electrons. This situation is known as *fermion sign problem* in the quantum Monte Carlo literature[1-4]. The solution of this problem is one of the most outstanding in all of the computational physics/chemistry. This problem is often (mis)understood as a technical detail that defeating the numerical simulators. To the best of our knowledge no methodology is available to handle this problem in a systematic and controlled fashion. However, we think that it is essentially a problem of exact boundary, which is not known for many electron systems for obtaining well-behaved solutions of non-relativistic Schrödinger equation. It is our understanding that the boundary must be derived from the link between the formal mathematics and the physics of the real world.

In this article, we will derive the boundary condition for atomic and molecular systems to obtain well-behaved solutions (i.e. bound state solution is single valued, continuous, quadratically integrable, and differentiable) of non-relativistic electronic Schrödinger equation. To start with, we are dealing with situations in which the ground state is non-degenerate only.

# THE EXACT BOUNDARY CONDITION

We have the time independent Schrödinger equation:

$$\hat{H}\,\Psi_0 = E_0\,\Psi_0 \qquad \ldots\ldots\ldots\ldots..(1)$$

where $\hat{H}$ is the time independent non-relativistic electronic Hamiltonian operator in the Born-Oppenheimer approximation, $E_0$ is the eigenvalue of the full many-electron ground state $\Psi_0$. The $\hat{H}$ is defined in atomic units as follows:

$$\hat{H} = -\frac{1}{2}\sum_{i=1}^{electrons}\nabla_i^2 - \sum_{i=1}^{electrons}V(\vec{r}_i) + \sum_{\substack{i=1\\j>i}}^{electrons}\frac{1}{r_{ij}} \qquad \ldots\ldots\ldots..(2)$$

where the external potential, $V(\vec{r}_i) = \sum_{I}^{Nuclei}\frac{Z_I}{r_{Ii}},$ \qquad \ldots\ldots\ldots.(3)

$\nabla^2$ is Laplacian, $Z_I$ denotes nuclear charge, and $r_{Ii}$ and $r_{ij}$ symbolize the electron-nucleus and electron-electron distance, respectively.

Following Hohenberg-Kohn theorem I, a proof only of existence[5], the electron density $\rho_0(\vec{r})$ in the ground state $\Psi_0$ is a functional of $V(\vec{r})$, i.e.

$$\rho_0(\vec{r}) = \rho_0[V(\vec{r})]. \qquad \ldots\ldots\ldots\ldots\ldots\ldots\ldots. (4)$$

Further, the full many electron ground state $\Psi_0$ is unique functional of $\rho_0(\vec{r})$, i.e.

$$\Psi_0 = \Psi_0[\rho_0(\vec{r})]. \qquad \ldots\ldots\ldots\ldots\ldots\ldots(5)$$

Evidently we can say that $\Psi_0$ is a functional of $V(\vec{r})$ i.e.

$$\Psi_0 = \Psi_0[V(\vec{r})]. \qquad \ldots\ldots\ldots\ldots\ldots (6)$$

We have a choice to express the exact density:

$$\rho_0[V(\vec{r})] = \sum_{i=1}^{N} \phi_i^2[V(\vec{r})] \quad \ldots\ldots\ldots\ldots(7)$$

Where $N$ denotes number of electrons. The functionals, $\{\phi_i[V(\vec{r})], i=1.....N\}$ are exact ortho-normal one electron functions of the function $V(\vec{r})$, which give exact $\rho_0[V(\vec{r})]$.

(Caution to reader!! At the moment, here is nothing to do with so-called *s, p, d, f,* ..etc. type orbitals. The functionals $\{\phi_i[V(\vec{r})], i=1.....N\}$ are entirely different from those orbitals obtained from Kohn-Sham[6] or similar formalisms.)

Now we can write the exact $N$ electron ground state wave function as a functional of $N$ exact one electron functionals $\{\phi_i[V(\vec{r})], i=1.....N\}$:

$$\Psi_0 = \Psi_0[\phi_1^2[V(\vec{r}_1)], \phi_2^2[V(\vec{r}_2)], ......, \phi_N^2[V(\vec{r}_N)]] \quad \ldots\ldots(8)$$

or $\Psi_0 = \Psi_0[\phi_1[V(\vec{r}_1)], \phi_2[V(\vec{r}_2)], ......, \phi_N[V(\vec{r}_N)]] \quad \ldots\ldots..(9)$

Since each one electron functional in $\{\phi_i[V(\vec{r})], i=1.....N\}$ is a function of external potential $V(\vec{r})$, we can also write the exact $N$ electron ground state wave function in functional form as follows:

$$\Psi_0 = \Psi_0[V(\vec{r}_1), V(\vec{r}_2), ......, V(\vec{r}_N)] \quad \ldots\ldots\ldots\ldots...(10)$$

Thus the exact $N$ electron non-degenerate ground state wave function is a unique functional of external potential experienced by each electron *i.e.* functional of $V(\vec{r}_1), V(\vec{r}_2), ......, V(\vec{r}_N)$.

So far, it is not clear:
- Whether the wave function is symmetric or antisymmetric with respect to interchange of any two electrons.
- What are the analytical forms of $\{\phi_i[V(\vec{r})], i=1.....N\}$?
- What is the analytical form of the exact wave function?

- How to get the exact wave function from the exact density.

However, we get some idea about the topology of a well behaved ground state non-degenerate wave function and distribution functions in a given external potential $V(\vec{r})$.

In particular: "*The probability of n-electrons (where n = 2..N) being found simultaneously on the isopotential surface of an external potential $V(\vec{r})$ is same irrespective of positions of the electrons on the surface.*"

Now we proceed to decide the nature (symmetric or antisymmetric with respect to interchange of any two electrons) of a well behaved many electron wave function.

Defining the local energy, $E_L$:

$$E_L = \frac{\hat{H}\Psi_0}{\Psi_0}$$

$$= -\frac{1}{2}\sum_{i=1}^{electrons}\frac{\nabla_i^2\Psi_0}{\Psi_0} - \sum_{i=1}^{electrons}\sum_{I}^{Nuclei}\frac{Z_I}{r_{Ii}} + \sum_{\substack{i=1\\j>i}}^{electrons}\frac{1}{r_{ij}} \quad\ldots\ldots\ldots\ldots(11)$$

The terms $Z_I/r_{Ii}$ and $1/r_{ij}$ in the equation (11) will blow up if $r_{Ii} \to 0$ and $r_{ij} \to 0$ unless so-called cusp conditions are obeyed by $\Psi_0$. The $\Psi_0$ is exact and obeys electron nucleus (*e-N*) and electron-electron (*e-e*) cusp conditions.

The wave function for a system of *N* identical particles must be symmetric or antisymmetric with regard to interchange of any two of the identical particles, *i* and *j*. Since the *N* particles are all identical, we could not have the wave function symmetric with regard to some interchanges and antisymmetric with regard to other interchanges. Thus the wave function of *N* identical particles must be either symmetric or antisymmetric with regard to every possible interchange of any two particles.

Let us assume $\Psi_0$ is symmetric with regard to interchange of electrons *i* and *j*. There is a cusp in $\Psi_0$ at $r_{ij} = 0$. This implies that $\Psi_0$ is not differentiable at $r_{ij} = 0$. Therefore, $\Psi_0$, symmetric with respect to interchange of any two electrons is not a well-behaved solution. To make $\Psi_0$ a well-behaved wave function, $\Psi_0$ must be zero when $r_{ij} = 0$ and also it must change sign with respect to the interchange of two electrons, i.e. if $\vec{r}_i = \vec{r}_j$ then $\Psi_0 = 0$. This condition is universal and independent of kind of external potential. However, we are interested in a well-behaved solution of a bound state in a given external potential $V(\vec{r})$. From the previous argument, we know that the simultaneous probability of finding two electrons is same everywhere at the isopotential surface. Therefore, if $V(\vec{r}_i) - V(\vec{r}_j) = 0$ then $\Psi_0 = 0$.

Extending to *N* electron system, we have

If $f = \prod_{\substack{i=1 \\ j>i}}^{N} \left( V(\vec{r}_i) - V(\vec{r}_j) \right) = 0$ then $\Psi_0 = 0$.

We can also express $f$ as Vandermonde determinant:

$$f = \prod_{\substack{i=1 \\ j>i}}^{N} \left( V(\vec{r}_i) - V(\vec{r}_j) \right) = \begin{vmatrix} 1 & 1 & \cdots & 1 & 1 \\ V(\vec{r}_1) & V(\vec{r}_2) & \cdots & V(\vec{r}_{N-1}) & V(\vec{r}_N) \\ V^2(\vec{r}_1) & V^2(\vec{r}_2) & \cdots & V^2(\vec{r}_{N-1}) & V^2(\vec{r}_N) \\ \cdot & \cdot & \cdots & \cdot & \cdot \\ \cdot & \cdot & \cdots & \cdot & \cdot \\ V^{N-1}(\vec{r}_1) & V^{N-1}(\vec{r}_2) & \cdots & V^{N-1}(\vec{r}_{N-1}) & V^{N-1}(\vec{r}_N) \end{vmatrix} \quad \ldots(12)$$

Consequently we have exclusion principle in the following form:

*"Two electrons can't be at the same external isopotential surface simultaneously."*

We see that if we are interested in a well behaved solution of the time independent Schrödinger equation, the boundary condition (12) (i.e. antisymmetric wave function) is obtained naturally due to singularity in the *e-e* interaction potential, which respects Pauli's exclusion principle. If electrons *i* and *j* are of opposite spin then we say that $V(\vec{r}_i) - V(\vec{r}_j) = 0$ represents Coulomb (nodal) surface. If electrons $i$ and $j$ are of same spin then $V(\vec{r}_i) - V(\vec{r}_j) = 0$ represents Coulomb-Exchange nodal surface. All together, the $f = \prod_{\substack{i=1 \\ j>i}}^{N} (V(\vec{r}_i) - V(\vec{r}_j)) = 0$ represents the Coulomb-Exchange nodal surface of $N$ electron system. Hereafter, we will call $f$ as $f_{Coulomb-Exchange}$ nodal surface. However, the solution obtained for the Hamiltonian (2) within the boundary $f_{Coulomb-Exchange} = 0$ does not tell us about the spin multiplicity of the $N$ electron system.

Further, we can rewrite the functional $f$ in terms of Hermite polynomials, $H_k[V(\vec{r})]$:

$$f[V(\vec{r})] = \begin{vmatrix} H_0[V(\vec{r}_1)] & H_0[V(\vec{r}_2)] & \cdots & H_0[V(\vec{r}_{N-1})] & H_0[V(\vec{r}_N)] \\ H_1[V(\vec{r}_1)] & H_1[V(\vec{r}_2)] & \cdots & H_1[V(\vec{r}_{N-1})] & H_1[V(\vec{r}_N)] \\ H_2[V(\vec{r}_1)] & H_2[V(\vec{r}_2)] & \cdots & H_2[V(\vec{r}_{N-1})] & H_2[V(\vec{r}_N)] \\ \cdot & \cdot & \cdots & \cdot & \cdot \\ \cdot & \cdot & \cdots & \cdot & \cdot \\ H_{N-1}[V(\vec{r}_1)] & H_{N-1}[V(\vec{r}_2)] & \cdots & H_{N-1}[V(\vec{r}_{N-1})] & H_{N-1}[V(\vec{r}_N)] \end{vmatrix} \quad \ldots(13)$$

In particular, if we multiply an optimizable one electron functional $\psi[V(r)]$ to the equation (13) and we obtain an $N$ electron wave function:

$$\Psi[V(\vec{r})] = Norm\ \psi[V(\vec{r}_1)]\psi[V(\vec{r}_2)]\ldots\psi[V(\vec{r}_N)]f[V(\vec{r})] \quad \ldots(14)$$

The one-electron density functional corresponds to the wave function (14):

$$\rho[V(\vec{r}_1);V(\vec{r}_1')] = \psi[V(\vec{r}_1)]\psi[V(\vec{r}_1')]\left(\sum_{k=0}^{N-1} A_k^2 H_k[V(\vec{r}_1)]H_k[V(\vec{r}_1')]\right) \quad \ldots(15)$$

where $A_k$ is normalization constant of $\psi[V(\vec{r})]H_k[V(\vec{r})]$.

The two-electron density functional in terms of one-electron density functional:

$$\begin{aligned}\pi[V(\vec{r}_1),V(\vec{r}_2);V(\vec{r}_1'),V(\vec{r}_2')] \\ = \rho[V(\vec{r}_1);V(\vec{r}_1')]\rho[V(\vec{r}_2);V(\vec{r}_2')] - \rho[V(\vec{r}_2);V(\vec{r}_1')]\rho[V(\vec{r}_1);V(\vec{r}_2')]\end{aligned} \quad \ldots(16)$$

Here it appears that we can obtain exact ground state energy by optimizing only a one-electron functional $\psi[V(r)]$ in the equation (14).

A very interesting and new physics obtained from the equation (13) is that each row in the determinant represents different level ($k$) of *Kamalpur breathing* (anharmonic quantum breathing) of electron cloud in a given external potential $V(\vec{r})$ and each level, $k$ is occupied by one electron (the elementary particle).

## BYPASSING THE SIGN PROBLEM

We can bypass the fermion sign problem in the electronic structure diffusion Monte Carlo (DMC) method using fixed node approach. Here one assumes a prior knowledge of the nodal surface i.e. $\Psi_0(R) = 0$. Due to tiling property[7] of the exact ground state wave function, the Schrödinger equation is solved in the volume embraced by the nodal surface, where the wave function has a constant sign and in this way the fermion sign problem is bypassed. The exact knowledge of Coulomb Exchange nodal surface allows us for an exact stochastic solution of the Schrödinger equation. The restriction in the random walk $R \rightarrow R'$ during the electronic structure diffusion Monte Carlo simulation is as follows:

$$\text{if } f_{Coulomb-Exchange}(R) f_{Coulomb-Exchange}(R') \begin{cases} > 0 & accept \\ < 0 & reject \end{cases} \quad \ldots\ldots\ldots\ldots(17)$$

We have applied the boundary condition (17) for the ground state electronic structure diffusion Monte Carlo simulation of N, Ne, $Li_2$, $Be_2$, $B_2$, $C_2$, $N_2$, $O_2$, $F_2$, and $H_2O$ systems.

Monte Carlo calculations can be carried out using sets of random points picked from any arbitrary probability distribution. The choice of distribution obviously makes a

---

difference to the efficiency of the method. If the Monte Carlo calculations are carried out using uniform probability distributions, very poor estimates of high-dimensional integrals are obtained, which is not a useful method of approximation. These problems are handled by introducing the importance sampling approach[8,9]. In this approach the sampling points are chosen from a trial distribution, which concentrates on points where the trial function, $\Phi_T(\mathbf{R})$ is large.

In the present DMC calculations, we have chosen the trial function, $\Phi_T$ in the form:

$$\Phi_T = \Phi.F, \qquad \qquad ....(18)$$

where $\Phi$ denotes the Hartree Fock (HF) or multi configuration self consistent field (MCSCF) wave function and $F$ is a correlation function that depends on inter-particle distances. The HF and MCSCF wave functions were obtained using the GAMESS package[10] and employed Dunning's *cc*-VTZ atomic basis set [11]. In order to satisfy the electron nucleus (*e-N*) cusp condition, all Gaussian type *s* basis functions were replaced with eight Slater-type *s* basis functions. The exponents of Slater-type *s* functions were taken from Koga *et al.* [12] and satisfy the *e-N* cusp condition.

We have chosen the Schmidt, Moskowitz, Boys, and Handy (SMBH) correlation function $F_{SMBH}$[13]. For the SMBH correlation function, Eqn. (19), we have included terms up to 2$^{nd}$ order, where order, *s* is defined as $s = l + m + n$.

$$F_{SMBH} = \exp\left(-\sum_{\mu}\sum_{A}^{atoms} c_{\mu A} \sum_{i>j}^{electrons} \left(\bar{r}_{iA}^{l_\mu} \bar{r}_{jA}^{m_\mu} + \bar{r}_{iA}^{m_\mu} \bar{r}_{jA}^{l_\mu}\right)\bar{r}_{ij}^{n_\mu}\right) \qquad ............(19)$$

where $\bar{r} = \dfrac{r}{1 + b\,r}$ \qquad\qquad ............(20)

and *r* denotes inter-particle distance. Six non-redundant parameters out of the total ten were optimized keeping $b = 1.0$ as follows:

1) First we obtained optimal parameters by minimizing the energy and variance at the variational Monte Carlo (VMC) level.
2) Using this VMC optimal trial function, the trial function fixed node DMC calculation was carried out and walkers were collected after each 2000 steps. Further, correlation parameters were reoptimized to minimize the variance with ~100,000 walkers. Here reference energy was set to the trial function fixed node DMC energy.

These optimized trial functions were used for importance sampling in the DMC simulation and a random walk $R \rightarrow R'$ was accepted if

$$f_{Coulomb-Exchange}(R) f_{Coulomb-Exchange}(R') > 0.$$

The DMC calculations were performed using the open source quantum Monte Carlo program, ZORI[14]. Around 10,000 walkers were used for the systems studied. The Umrigar et al.[15] algorithm was used for DMC walks and Caffarel Assaraf et al.[16] algorithm for population control. We have allowed only one electron walk at a time. The DMC calculations were done at several time steps. We report only those energies extrapolated to zero time step.

We present the ground state DMC energies of N, Ne, $Li_2$, $Be_2$, $B_2$, $C_2$, $N_2$, $O_2$, $F_2$, and $H_2O$ systems in Table I. The DMC energies obtained using our newly derived boundary $f_{Coulomb-Exchange} = 0$ are far better than the trial function fixed node DMC energies[17] and compare well with the experimental counterpart. However, present simulations were noisy and unpleasant compared to conventional trial function fixed node DMC simulations. It is worth noting that we have obtained DMC energy even lower than the exact value at smaller time steps for the atoms of relatively larger atomic radius perhaps due to failure of the distributions to reach the steady state or equilibrium distributions in a finite number of steps. This problem can be handled in the Green's function quantum Monte Carlo (GFQMC) method as it takes the advantage of the properties of Green's functions in eliminating time-step entirely in treating the steady state equation. The GFQMC is well suited if boundaries are exactly known[18]. If the trial function boundary and the $f_{Coulomb-Exchange} = 0$ does not coincide and also non-zero values of trial function are very much different from the exact solution, which could lead to large statistical fluctuations from poor sampling and possibly to an effective non-ergodic diffusion process due to the finite projection time in practical calculations. Therefore, we are looking for an alternative well behaved trial function whose boundary coincides with those of $f_{Coulomb-Exchange}$.

## CONCLUSION

This article presents a progress of the author's research in order to get exact solution of non-relativistic Schrödinger equation of many electron systems. A conclusion of this on going research is that we have derived the exclusion principle "Two electrons can't be at the same external isopotential surface simultaneously" using the first postulate of quantum mechanics. We propose the exact Coulomb-Exchange nodal surface i.e. the exact boundary to solve the non-relativistic Schrödinger equation for non-degenerate ground state of atoms and molecules. Using this newly derived boundary condition, one can bypass the fermion sign problem in the electronic structure Quantum Monte Carlo simulation and hence the exact ground state energy as well as the exact electron density.

# ACKNOWLEDGEMENTS

The QMC calculations were carried out at the Lawrence Berkeley National Laboratory, Berkeley. The author gratefully acknowledges Professor W. A. Lester for his support during the stay at Berkeley. The author is indebted to Professor P. Chandra of Banaras Hindu University, Varanasi for his interest and helpful discussion during the preparation of the manuscript. Professor S. K. Sengupta of Banaras Hindu University, Varanasi is acknowledged for careful reading of the manuscript.

TABLE I. The total ground state energies obtained from fixed-node DMC calculation.

| Atom / Molecule | Bond length | CSF,D | $E_{TFN-DMC}$ (Ref. 17) | $E_{CEN-DMC}$ (Extrapolated to $\tau=0$) | $E_0$ |
|---|---|---|---|---|---|
| N |  | 1,1 | -54.5753(3) | **-54.5902(11)** | *-54.5892* |
|  |  | -, 111 | -54.5841(5) |  |  |
| Ne |  | 1,1 | -128.9216(15) | **-128.938(1)** | *-128.9375* |
| Li$_2$ | 5.051 | 1,1 | -14.9911(1) | **-14.9955(5)** | *-14.9954* |
|  |  | 4,5 | -14.9938(1) |  |  |
| Be$_2$ | 4.63 | 1,1 | -29.3176(4) | **-29.3378(15)** | *-29.33854(5)* |
|  |  | 5,16 | -29.3301(2) |  |  |
| B$_2$ | 3.005 | 1,1 | -49.3778(8) | **-49.41655(45)** | *-49.415(2)* |
|  |  | 6,11 | -49.3979(6) |  |  |
| C$_2$ | 2.3481 | 1,1 | -75.8613(8) | **-75.9229(19)** | *-75.923(5)* |
|  |  | 4,16 | -75.8901(7) |  |  |
|  |  | **77, 314** | **-75.9035(9)** |  |  |
| N$_2$ | 2.068 | 1,1 | -109.487(1) | **-109.5424(15)** | *-109.5423* |
|  |  | 4,17 | -109.505(1) |  |  |
|  |  | -, 552 | -109.520(3) |  |  |
| O$_2$ | 2.282 | 1,1 | -150.268(1) | **-150.3274(15)** | *-150.3268* |
|  |  | 4,7 | -150.277(1) |  |  |
| F$_2$ | 2.68 | 1,1 | -199.478(2) | **-199.5289(25)** | *-199.5299* |
|  |  | 2,2 | -199.487(1) |  | *-199.52891(4)* |
| H$_2$O |  | 1,1 | -76.4175(4) | **-76.4376(11)** | *-76.438(3)* |
|  |  | -, 300 | -76.429(1) |  | *-76.4376* |

Bond lengths and energies are in atomic units. In the third column, we list the number of configuration state functions (CSFs) and number of determinants (D) in the trial function ($\Phi_T$). $E_{TFN-DMC}$ denotes the DMC energy with $\Phi_T = 0$. $E_{CEN-DMC}$ denotes the DMC energy with $f_{Coulomb-Exchange} = 0$. $E_0$ denotes the exact, non-relativistic, infinite nuclear mass energy. The numbers shown in bracket are error bar.

# Supplementary note for reviewers:

1. The proposed theory is to deal only real interacting many electron systems. To start with only non-degenerate ground state of many electron systems are considered. Author is neither interested nor intended to deal any kind of non-interacting systems such as free fermion, free electron gases, or free particles because author think that none of the real system belong to either of these classes. Author has chosen to construct the boundary condition from the link between the formal mathematics and the physics of many electron systems.

2. **Difference between spatial nodes and Coulomb-Exchange nodal surface:**
   I hope that people can distinguish spatial nodes and Coulomb Exchange nodes and the physics behind the different kind of nodes. Whatever I have discussed in this paper is only about Coulomb-Exchange nodal surfaces. There is no analogy with a particle in a box node and Coulomb-Exchange nodal surfaces. For example: The function $f(r_1,r_2)=(r_1-1)(r_2-1)(r_1-r_2)\exp(-2r_1-2r_2)$ is antisymmetric with respect to interchange of two electrons. However, the node $(r_1-1)(r_2-1)$ is symmetric with respect to interchange of two electrons and fixed and this node can be compared with nodes of particle in a box. The Coulomb-Exchange node $(r_1-r_2)$ is antisymmetric with respect to interchange of two electrons and responsible for removal of singularity in e-e interaction potential. The Coulomb-Exchange nodal surfaces only occur in a system of more than one electron systems. Author understands that the Coulomb-Exchange nodal surfaces are directly responsible for the existence of real many electron systems.

3. **A consequence of proposed solution of the sign problem is that the ground state of Helium atom in the non-relativistic limit has a nodal surface. However, it is understood that He ground state wave function is symmetric and has no such node.**

A consequence of proposed solution for the sign problem is that the ground state of the Helium atom in the non-relativistic limit has **Coulomb-Exchange nodal surface, $r_1$-$r_2$=0**.

An understanding that He atom ground state wave function is symmetric and has no such node is an *illusion* only. This illusion arises due to a practice that the QMC people using phi(1)*phi(2)*Jastrow trial function, where phi(r) is 1s orbital. The trial function is symmetric with respect to exchange of two electrons. The trial function also satisfies electron nucleus cusp condition. We also expect that the final solution will satisfy *e-e* cusp condition. *Since the trial function is symmetric, people got accurate energy and assumed that the final solution is also symmetric and it does not have any node also wave function is non-zero at the point of coincidence of two electrons*. Where is Coulomb hole? However, it can be proven that a symmetric solution is not acceptable. Proofs are as follows:

## "Proof for the existence of Coulomb-Exchange node in He ground state exact wave function"

**A.)** Let assume

$\Psi_{sym}(\vec{r}_1, \vec{r}_2)$ is an exact symmetric wave function.

i.e. $\Psi_{sym}(\vec{r}_1, \vec{r}_2) = \Psi_{sym}(\vec{r}_2, \vec{r}_1)$

Since $\Psi_{sym}(\vec{r}_1, \vec{r}_2)$ is exact, it must satisfy the cusp condition at $\vec{r}_1 = \vec{r}_2$. Clearly there is a cusp at $\vec{r}_1 = \vec{r}_2$.

Since there is a cusp at $\vec{r}_1 = \vec{r}_2$ in $\Psi_{sym}(\vec{r}_1, \vec{r}_2)$, the second derivative

$\partial^2 \Psi_{sym}(\vec{r}_1, \vec{r}_2) / \partial x_1^2$ is not defined at $\vec{r}_1 = \vec{r}_2$.

Therefore $\Psi_{sym}(\vec{r}_1, \vec{r}_2)$ is not a well behaved solution and hence it is not an acceptable wave function. Only option left is antisymmetric solution.

**B.)** Another proof:

$$\Psi(\vec{r}_1,\vec{r}_2) = \sum_{\mu}^{\infty}\sum_{\nu}^{\infty} c_{\mu\nu}\phi_{\mu}(\vec{r}_1)\phi_{\nu}(\vec{r}_2) \qquad \ldots\ldots\ldots..(S-1)$$

Where $\{\phi_{\mu}(\vec{r})\}$ forms a real infinite one-electron basis.

Since $\Psi(\vec{r}_1,\vec{r}_2)$ is expanded over infinite basis set and hence it is exact.

This implies that $\Psi(\vec{r}_1,\vec{r}_2)$ satisfies the cusp condition at $\vec{r}_1 = \vec{r}_2$.

$$\nabla_1^2\Psi(\vec{r}_1,\vec{r}_2) = \sum_{\mu}^{\infty}\sum_{\nu}^{\infty} c_{\mu\nu}\nabla_1^2\phi_{\mu}(\vec{r}_1)\phi_{\nu}(\vec{r}_2). \qquad \ldots\ldots\ldots..(S-2)$$

The second derivative $\nabla_1^2\Psi(\vec{r}_1,\vec{r}_2)$ is continuous at $\vec{r}_1 = \vec{r}_2$ because each term in the expansion is continuous (the rules of continuity for algebraic combinations).

This implies that there is no cusp in $\Psi(\vec{r}_1,\vec{r}_2)$ at $\vec{r}_1 = \vec{r}_2$.

BUT $\Psi(\vec{r}_1,\vec{r}_2)$ has to satisfy the cusp condition at $\vec{r}_1 = \vec{r}_2$.

This is only possible if $\Psi(\vec{r}_1,\vec{r}_2)$ changes the sign at $\vec{r}_1 = \vec{r}_2$.

And hence the exact $\Psi(\vec{r}_1,\vec{r}_2)$ has exchange node irrespective of its spin multiplicity.

**C.)** More illustrative example:

Hamiltonian for He atom:

$$\hat{H} = -\frac{1}{2}\nabla_1^2 - \frac{1}{2}\nabla_2^2 - \frac{2}{r_1} - \frac{2}{r_2} + \frac{1}{r_{12}} \qquad \ldots\ldots\ldots..(S-3)$$

and $H\Psi(1,2) = E\Psi(1,2)$ $\qquad \ldots\ldots\ldots..(S-4)$

Let expand

$$\Psi(1,2) = \sum_{\mu}^{\infty}\sum_{\nu}^{\infty} c_{\mu\nu}\phi_{\mu}(1)\phi_{\nu}(2) \qquad \ldots\ldots\ldots..(S-5)$$

Where $\{\phi_{\mu}(r)\}$ is complete set of eigen functions of the Hamiltonian

$\hat{h} = -\frac{1}{2}\nabla^2 - \frac{2}{r}$ with eigenvalue equation $\hat{h}\phi_{\mu}(r) = \varepsilon_{\mu}\phi_{\mu}(r)$.

Rewriting the He Hamiltonian:

$$\hat{H} = -\frac{1}{2}\nabla_1^2 - \frac{1}{2}\nabla_2^2 - \frac{2}{r_1} - \frac{2}{r_2} + \frac{1}{r_{12}} = \hat{h}_1 + \hat{h}_2 + \frac{1}{r_{12}} \quad \ldots\ldots(S\text{-}6)$$

$$E_L = \frac{\hat{H}\Psi(1,2)}{\Psi(1,2)} = \frac{(\hat{h}_1 + \hat{h}_2 + \frac{1}{r_{12}})\Psi(1,2)}{\Psi(1,2)} = \frac{(\hat{h}_1 + \hat{h}_2)\sum_{\mu}^{\infty}\sum_{\nu}^{\infty}c_{\mu\nu}\phi_\mu(1)\phi_\nu(2)}{\Psi(1,2)} + \frac{1}{r_{12}}$$
$$\ldots\ldots(S\text{-}7)$$

Since $\phi_\mu(r)$ is an eigen function of $\hat{h}$.

We can write

$$E_L = \frac{\sum_{\mu}^{\infty}\sum_{\nu}^{\infty}c_{\mu\nu}(\varepsilon_\mu + \varepsilon_\nu)\phi_\mu(1)\phi_\nu(2)}{\Psi(1,2)} + \frac{1}{r_{12}} \quad \ldots\ldots(S\text{-}8)$$

$$E = \varepsilon_\mu + \varepsilon_\nu + d_{\mu\nu}$$

$$\varepsilon_\mu + \varepsilon_\nu = E - d_{\mu\nu} \quad \ldots\ldots(S\text{-}9)$$

$$E_L = \frac{\sum_{\mu}^{\infty}\sum_{\nu}^{\infty}c_{\mu\nu}(E - d_{\mu\nu})\phi_\mu(1)\phi_\nu(2)}{\Psi(1,2)} + \frac{1}{r_{12}} \quad \ldots\ldots(S\text{-}10)$$

$$E_L = \frac{E\sum_{\mu}^{\infty}\sum_{\nu}^{\infty}c_{\mu\nu}\phi_\mu(1)\phi_\nu(2)}{\Psi(1,2)} - \frac{\sum_{\mu}^{\infty}\sum_{\nu}^{\infty}c_{\mu\nu}d_{\mu\nu}\phi_\mu(1)\phi_\nu(2)}{\Psi(1,2)} + \frac{1}{r_{12}} \quad \ldots\ldots(S\text{-}11)$$

$$E_L = \frac{E\Psi(1,2)}{\Psi(1,2)} - \frac{\sum_{\mu}^{\infty}\sum_{\nu}^{\infty}c_{\mu\nu}d_{\mu\nu}\phi_\mu(1)\phi_\nu(2)}{\Psi(1,2)} + \frac{1}{r_{12}} \quad \ldots\ldots(S\text{-}12)$$

$$E_L = E - \frac{\sum_{\mu}^{\infty}\sum_{\nu}^{\infty}c_{\mu\nu}d_{\mu\nu}\phi_\mu(1)\phi_\nu(2)}{\Psi(1,2)} + \frac{1}{r_{12}} \quad \ldots\ldots(S\text{-}13)$$

If $\Psi(1,2)$ is exact then

$E_L = E$

and

$$-\frac{\sum_{\mu}^{\infty}\sum_{\nu}^{\infty} c_{\mu\nu} d_{\mu\nu} \phi_\mu(1)\phi_\nu(2)}{\Psi(1,2)} + \frac{1}{r_{12}} = 0 \qquad \ldots\ldots\ldots\text{(S-14)}$$

Now let assume that the exact solution is symmetric with respect to interchange of two electrons. $\Psi(1,2)$ and $\phi_\mu(r)$ are well behaved and differentiable. From the rules of continuity for algebraic combinations, the term in equation (S-14), $\dfrac{\sum_{\mu}^{\infty}\sum_{\nu}^{\infty} c_{\mu\nu} d_{\mu\nu} \phi_\mu(1)\phi_\nu(2)}{\Psi(1,2)}$ is continuous and finite and it should not diverge when $r_{12} \to 0$. *Therefore, symmetric solution is not acceptable*.

However, if $\Psi(1,2) = 0$ at $r_{12}=0$ then $\dfrac{\sum_{\mu}^{\infty}\sum_{\nu}^{\infty} c_{\mu\nu} d_{\mu\nu} \phi_\mu(1)\phi_\nu(2)}{\Psi(1,2)}$ will also diverge and can compensate the divergence in $1/r_{12}$ term.

***Thus the only acceptable solution is antisymmetric (with respect to interchange of two electrons) solution.***

**D.) Another example:**

Almost all QMC people believe (their believe is based on some assumptions and approximations) that He atom ground state wave function is symmetric. This is an *illusion*. This can be understood as follows:

Let us take trial functions of two electron system:

$$U = \frac{a\sqrt{(x_1 - x_2)^2}}{1 + b\sqrt{(x_1 - x_2)^2}}$$

$$g_1(x_1, x_2) = e^{-2x_1} e^{-2x_2} e^{U}$$

$$g_2(x_1,x_2) = \sqrt{(x_1-x_2)^2}\ e^{-2x_1} e^{-2x_2} e^U$$

$$g_3(x_1,x_2) = (x_1-x_2)e^{-2x_1} e^{-2x_2} e^U$$

If someone claims that He ground state is symmetric, what kind of exact symmetric wave functions they are getting finally? The functions $g_1$ and $g_2$ are symmetric with respect to interchange of two electrons. The functions like $g_1$, $g_2$, and $g_3$ can satisfy cusp condition. The functions $g_1$ and $g_2$ are not differentiable at $x_1=x_2$ and therefore these are not acceptable. The antisymmetric function $g_3$ are differentiable at $x_1=x_2$.

However, people have got very accurate ground state energy for He atom using HF*Jastrow trial function and they concluded that He atom has no node without examining the simultaneous probability of finding two electrons at exactly same place. I think, they got good results due to inherent beauty of DMC technique.

The functions $g_2$ is symmetric and $g_3$ is antisymmetric with respect to interchange of two electrons. However, $g_2*g_2$ and $g_3*g_3$ give the exactly same probability distribution i.e. same physics. Functions $g_2$ and $g_3$ vanish when $x_1=x_2$. Further, the VMC calculation for $g_2$ and $g_3$ will give exactly same energy. Can anyone predict that the VMC energies obtained from $g_2$ and $g_3$ represent singlet or triplet state? I am sure it is not possible.

An antisymmetric wave function can satisfy cusp condition as well as it's derivative will be continuous simultaneously at the point of coincidence. Here symmetric and antisymmetric wave functions serve same distribution. Why I should not prefer antisymmetric wave function for which a boundary condition can be imposed easily?

4. **Further, if I assume the argument "He ground state wave function is symmetric and has no such node" is correct. The end will be a nonsense, which is as follows:**

It is very much common practice in QMC calculation to take Hartree-Fock trial function as a product of alpha-beta determinants. For example N atom:

$PSI_T(1,2,3,4,5,6,7) = Det(1,2,3,4,5,6,7)$.

$PSI_T(1,2,3,4,5,6,7) = Det_\alpha(1,2,3,4,5)*Det_\beta(6,7)$.

$PSI_T(1,2,3,4,5,6,7) = Det_\alpha(1,2,3,4,5)*Det_\beta(6,7)$.

$Det_\alpha(1,2,3,4,5)*Det_\beta(6,7) \neq Det_\alpha(1,2,3,4,7)*Det_\beta(6,5)$

The trial function $PSI_T$ is clearly neither symmetric nor antisymmetric with respect to interchange of alpha-beta electrons. It is not clear to me what kind of final solution (i.e. symmetric or antisymmetric) we will obtain with this trial function fixed node DMC? The fact that we can not write $PSI_{exact} = Phi_\alpha * Phi_\beta$.

5. **It is natural to ask what is node of $^3S$ He atom and why people are getting very accurate energy with exchange node $r_1-r_2=0$?**

   At the moment, I can only say that this is due to artifact of importance sampling because people have used HF*Jastrow trial function. The correlation energy for He($^3S$) atom is around 2mH and the overlap of HF trial function with exact wave function can be anticipated to be more than 99%. Perhaps due to technical reasons final DMC solution may have converged to He($^3S$). I have seen some recent papers on the node of He($^3S$) system. It is widely claimed that the node $r_1-r_2=0$ belongs to He($^3S$) system and it is exact. I differ with their argument and I proved that the exchange node $r_1-r_2=0$ belongs to He ground state. I do not know if anyone have performed DMC calculation with a trial function like $psi_t(r_1,r_2)=(r_1-r_2)*exp(-2*r_1)*exp(-2*r_2)$ and reported the energy for He($^3S$). Anyway, at present I am interested only in the non-degenerate ground state of atoms and molecules.

**Author welcomes further comments, questions, and suggestions if any.**
**Theochem@gmail.com**